\documentclass[10pt, conference, letterpaper]{IEEEtran}
\IEEEoverridecommandlockouts
\usepackage{cite}
\usepackage{amsmath,amssymb,amsfonts}
\usepackage{multirow}
\usepackage{graphicx}
\usepackage{textcomp}
\usepackage{algorithm}
\usepackage{algpseudocode}
\usepackage{comment}
\usepackage{xcolor}
\usepackage[nowarn,acronyms,nonumberlist,nopostdot,nomain,nogroupskip]{glossaries}
\usepackage{pifont}

\usepackage{pgfplots}
\usepackage{makecell}
\usepackage{amsmath}

\usepackage[top=0.73in, left=0.645in, right=0.645in, bottom=1.045in]{geometry}

\newacronym{3gpp}{3GPP}{3rd Generation Partnership Project}
\newacronym{itu}{ITU}{International Telecommunication Union}
\newacronym{aerpaw}{AERPAW}{Aerial Experimentation and Research Platform for Advanced Wireless}
\newacronym{wifi}{WiFi}{Wireless Fidelity}
\newacronym{sls}{SLS}{System Level Simulations}
\newacronym{kpis}{KPIs}{Key Performance Indicators}
\newacronym{kpm}{KPM}{key performance metric}
\newacronym{lte}{LTE}{Long Term Evolution}
\newacronym{mpquic}{MPQUIC}{Multi-Path QUIC}
\newacronym{ap}{AP}{access point}
\newacronym{aps}{APs}{access points}
\newacronym{rssi}{RSSI}{received signal strength index}
\newacronym{bs}{BS}{base station}
\newacronym{vho}{VHO}{vertical handover}
\newacronym{ml}{ML}{machine learning}
\newacronym{lstm}{LSTM}{long short term memory}
\newacronym{qos}{QOS}{quality of service}
\newacronym{uma}{UMa}{urban macro}
\newacronym{inh}{InH}{indoor hall}
\newacronym{i2i}{I2I}{Indoor to Indoor}
\newacronym{i2o}{I2O}{Indoor to Outdoor}
\newacronym{o2i}{O2I}{Outdoor to Indoor}
\newacronym{o2o}{O2O}{Outdoor to Outdoor}
\newacronym{los}{LOS}{line-of-sight}
\newacronym{nlos}{NLOS}{non line of sight}
\newacronym{rsrp}{RSRP}{reference signal received power}
\newacronym{rsrq}{RSRQ}{reference signal received quality}
\newacronym{pci}{PCI}{physical cell ID}
\newacronym{ta}{TA}{timing advance}
\newacronym{bssid}{BSSID}{basic service set identifier}
\newacronym{ssid}{SSID}{service set identifier}
\newacronym{rl}{RL}{reinforcement learning}
\newacronym{dqn}{DQN}{deep Q-network}
\newacronym{gsm}{GSM}{Global System for Mobile}
\newacronym{nr}{NR}{5G New Radio}
\newacronym{hetnets}{HetNets}{Heterogeneous Networks}
\newacronym{qoe}{QOE}{quality of experience}
\newacronym{gps}{GPS}{Global Positioning System}
\newacronym{swcp}{SWCP}{sticky WiFi client problem}
\newacronym{iot}{IOT}{Internet of Things}
\newacronym{uav}{UAV}{Unmanned Aerial Vehicles}
\newacronym{oran}{ORAN}{Open Radio Access Networks}
\newacronym{rem}{REM}{Radio Environment Map}
\newacronym{rma}{RMa}{Rural Macro Environment}
\newacronym{lw}{LW}{Lake Wheeler}
\newacronym{rf}{RF}{Radio Frequency}
\newacronym{rdz}{RDZ}{Radio Dynamic Zone}
\newacronym{son}{SON}{Self-Organizing Networks}
\newacronym{nlp}{NLP}{Natural Language Processing}
\newacronym{fspl}{FSPL}{free space path loss}
\newacronym{lwsrd}{LWSRD}{linear warmup and square-root decay}
\newacronym{gat}{GAT}{graph attention network}
\newacronym{gan}{GAN}{generative adversarial network}
\newacronym{mae}{MAE}{mean absolute error}
\newacronym{rmse}{RMSE}{root mean square error}
\newacronym{gpr}{GPR}{Gaussian process regression}
\newacronym{mse}{MSE}{mean square error}

\def\BibTeX{{\rm B\kern-.05em{\sc i\kern-.025em b}\kern-.08em
    T\kern-.1667em\lower.7ex\hbox{E}\kern-.125emX}}
    
\begin{document}

\title{TransfoREM: Transformer aided 3D Radio Environment Mapping}

\author{\IEEEauthorblockN{Gautham Reddy*, Ismail G\"{u}ven\c{c}*, Mihail L. Sichitiu*, Arupjyoti Bhuyan\textsuperscript{\textdagger}, Bryton Petersen\textsuperscript{\textdagger}}
\IEEEauthorblockN{ Jason Abrahamson\textsuperscript{\textdagger}}
\IEEEauthorblockA{*Department of Electrical and Computer Engineering, NC State University, Raleigh, NC 27606}
\IEEEauthorblockA{\textsuperscript{\textdagger}Idaho National Laboratory, Idaho Falls, ID 83415}
Email:\{greddy2, iguvenc, mlsichit\}@ncsu.edu,\{arupjyoti.bhuyan, Bryton.Petersen, jason.abrahamson\}@inl.gov
\thanks{ We use measurement data from prior published works~\cite{SungJoonKriging,empirical3dchannelmodeling}. This research is supported in part by the NSF award CNS-2332835 and the INL Laboratory Directed Research Development (LDRD) Program under BMC No. 264247, Release No. 26 on BEA's Prime Contract No. DE-AC07-05ID14517. Corresponding author: Ismail G\"{u}ven\c{c} (email: iguvenc@ncsu.edu).}
}

\maketitle

\begin{abstract}
Providing reliable cellular connectivity to \gls{uav} is a key challenge, as existing terrestrial networks are deployed mainly for ground-level coverage. 
% Extending these networks skyward leads to a limited service range and unpredictable signal strength from antenna side lobes, with such problems further exacerbated by \gls{uav} flight dynamics.
The cellular network coverage may be available for a limited range from the antenna side lobes, with poor connectivity further exacerbated by \gls{uav} flight dynamics.
In this work, we propose TransfoREM, a 3D \gls{rem} generation method that combines deterministic channel models and real-world data to map terrestrial network coverage at higher altitudes. 
At the core of our solution is a transformer model that translates radio propagation mapping into a sequence prediction task to construct \gls{rem}s. 
Our results demonstrate that TransfoREM offers improved interpolation capability on real-world data compared against conventional Kriging and other \gls{ml} techniques. 
Furthermore, TransfoREM is designed for holistic integration into cellular networks at the \gls{bs} level, where it can build \gls{rem}s, which can then be leveraged for enhanced resource allocation, interference management, and spatial spectrum utilization.
\end{abstract} 

\begin{IEEEkeywords}
3D Radio Environment Maps, Transformers, Unmanned Aerial Vehicle. 
\end{IEEEkeywords}

\glsresetall

\section{Introduction}

The traditional approach to wireless network planning has relied on deterministic propagation models.
However, the future network landscape will be more dynamic, with complex interactions arising from extreme densification, support for a wider variety of applications, and the integration of new high-frequency bands.
Such advances will present challenges in interference management, enhanced coverage, and energy utilization, necessitating more data-driven and real-time network management tools.
To address these needs, future networks will benefit from the \gls{rem} capability to proactively maintain a spatio-temporal database of the \gls{rf} environment in the network vicinity~\cite{EngineeringREM}. 

\begin{figure*}
	\includegraphics[width=0.95\linewidth]{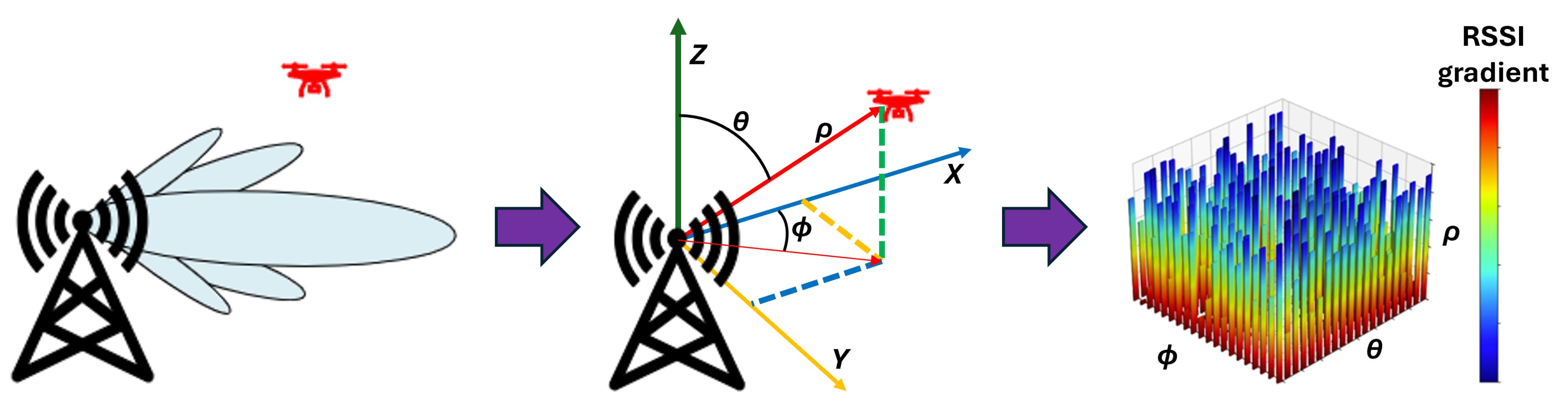}
	\centering
	\caption{The spherical coordinate system representation of a \gls{uav} position and its associated received signal strength indicator (RSSI) sequences in space.}
    \label{fig:Spherical_Coord}
\end{figure*}

\gls{rem}s serve as a comprehensive view of the network coverage over space, helping to solve key challenges such as:
\begin{itemize}
    \item \textbf{Dynamic Spectrum Access:} Wireless spectrum is a valuable and increasingly scarce resource. Existing practices of fixed-spectrum allocation lead to under-utilization, creating scope for sharing based on regional usage. \gls{rem}s provide a solution to this problem by acting as spectrum usage mapping tools, allowing networks to identify and utilize underused frequency bands across space. This capability augments the \gls{rdz}~\cite{SungJoonKriging} concept being developed for new users to harmoniously share spectrum with incumbent users.
    \item \textbf{Serving Aerial Users:} Terrestrial cellular networks, designed for ground-level users, pose coverage challenges for high-altitude applications~\cite{REM_connectedUAV}. Aerial users have a clear \gls{los} to multiple \gls{bs}s simultaneously, but they also experience varying antenna patterns due to the downward tilt of terrestrial \gls{bs}s. These effects lead to complicated coverage management and high-interference or weak signal zones in 3D space. Hence, a 3D \gls{rem} helps to understand and better manage connectivity at higher altitudes. 
    % \item \textbf{\gls{son}:} Traditional networks struggle to account for new signal sources and interference caused by the dense deployment of small cells. \gls{rem}s can alleviate these problems by collecting data from nearby users to map regional defects. A \gls{son} can then detect anomalies or reconfigure antenna beams to adapt to the changing environment~\cite{SiteAntennaMap}.
\end{itemize}
\gls{rem}s help wireless networks overcome these challenges, transforming them to optimize future performance proactively.
The development of effective \gls{rem} construction techniques, ranging from parametric models to deep learning-based spatial mapping~\cite{REM_Approaches}, has been essential to achieve this goal. 
Toward this end, we introduce TransfoREM to estimate the radio propagation by modeling signal strength as sequences originating from the \gls{bs} as seen in Fig.~\ref{fig:Spherical_Coord}.
This framework is specifically designed for deployment at individual \gls{bs}s, where it can learn the local radio footprint to manage better connectivity, interference, and mobility within its coverage area.
% In this work, we develop TransfoREM to estimate radio propagation behavior as signal strength sequences originating from the \gls{bs} . 
% It is designed for deployment at each \gls{bs} capable of learning its radio footprint to manage connectivity, interference, and mobility in its vicinity.  
The main contributions of this work are as follows. 
\begin{enumerate}
    \item Radio propagation modeling is posed as a sequence estimation task in the spherical coordinate system, aligning with the nature of outward radiation and antenna directivity gains inherent to wireless signal propagation. 
    \item We show that a transformer model with masked sequence translation, a technique borrowed from language processing, can accurately predict radio propagation. This method is effective for mapping radio environments with limited real-world data.
    \item Our approach demonstrates a true 3D \gls{rem} generation framework that extrapolates from both 2D sample measurement planes and 3D spatial data, outperforming Kriging and \gls{ml} techniques.
\end{enumerate} 

The rest of the paper is organized as follows. Section~\ref{sec:LitReview} provides a literature review of related work.
Section~\ref{sec:SystemModel} details the radio propagation system model, real-world data collection, and analysis. 
Section~\ref{sec:SystemDesign} then discusses the transformer model used to generate \gls{rem}s. 
In Section~\ref{sec:Results}, we present the experimental results, including comparative case studies using the data and results from~\cite{SungJoonKriging, MushfiqurKriging, empirical3dchannelmodeling}. 
Finally, Section~\ref{sec:Conclusion} concludes the article.

\section{Related Work}\label{sec:LitReview}

With \gls{rem}s holding significant importance in wireless network planning and operation, they have been extensively studied using various reconstruction techniques. 
Broadly, these \gls{rem} techniques can be categorized into model-driven methods, data-driven methods, and a hybrid of model- and data-driven methods~\cite{REM_survey}.  
The model-driven methods rely on path loss models, geometry-based ray tracing methods, and stochastic models that capture the behavior of received signal strength. 
However, such methods are limited by the practical deviations such as multi-path and diffraction effects in real-world.

Recent approaches, such as~\cite{REM_Empirical} have focused more on incorporating real-world data through data-driven \gls{rem} mapping techniques. 
A major hurdle for these techniques, however, is the high sampling requirement, as both 2D and 3D \gls{rem} reconstruction benefit from a dense sampling rate. 
Prior work has sought to address this by identifying and prioritizing specific spatial regions based on the environment's geometric and terrain information to focus data collection efforts~\cite{ActiveREM_UAV, TSVD_3DREM}.   
Recognizing the shortcomings of both stand-alone model and data-driven techniques, the hybrid (model + data-driven) \gls{rem} approach offers a robust solution by complementing the advantages of each. 
In this work, we utilize this hybrid methodology to develop a REM framework designed for deployment at individual BS sites, enabling the learning of the unique local radio propagation environment.

Recent hybrid \gls{rem} generation approaches include~\cite{RadioGAT}, where a \gls{gat} is leveraged to fuse model-based spatio-spectral correlation with data-driven radiomap generation. Similarly,~\cite{REM_ConstrainedUAV} employs a \gls{gan} with unsupervised learning to build \gls{rem}s from sparse signal strength measurements. 
Other notable methods include the triple-layer \gls{ml} technique in ~\cite{empirical3dchannelmodeling}, which uses a stagewise process of linear regression, ensemble methods, and a \gls{gpr} to predict and refine spatial radio \gls{kpis} using real data. 
Furthermore, \cite{STORM} introduces a transformer-based technique to learn spatial correlation trends, effectively mimicking the Kriging interpolation method without requiring a semi-variogram. 
While these methods are all novel in their approaches, our proposed work distinguishes itself as a truly physics-inspired technique that traces radio wave propagation by learning foundational model knowledge and then adapting according to measurement data. 

\section{System Model And Data Analysis}\label{sec:SystemModel}

Wireless channel modeling is a widely researched area that serves as a fundamental step in system design and analysis. 
Spatial channel modeling is a subset of this effort, with standardization bodies such as \gls{3gpp} publishing detailed channel characteristic models that range from large-scale outdoor channels to indoor environmental channels.  
Generally, these statistical channel models do not accurately capture the site-specific channel behavior and need to be augmented with data-driven perception to improve connectivity management at individual \gls{bs}s.
To build this capability, we begin with a propagation channel model abstraction similar to the one presented in~\cite{UAVcorridors}.

\subsection{Deterministic Channel Propagation Model}
For a user $k$ connected to a \gls{bs} $b$, the received signal strength $P_{b,k_{\mathrm{dB}}}(\rho,\phi,\theta)$ as a function of radial separation $\rho$ and 3D angles $\phi,\theta$ according to Fig.~\ref{fig:Spherical_Coord}, in dB-scale is defined as: 
\begin{equation}
    \begin{aligned}
    & P_{b,k_{\mathrm{dB}}}(\rho,\phi,\theta) = P_{b_\mathrm{dB}} + G_{b,k_\mathrm{dB}}(\rho,\phi,\theta) + N, \\
    \end{aligned}
\end{equation}
where $P_{b_\mathrm{dB}}$ is the transmit power of \gls{bs} $b$, $G_{b,k_\mathrm{dB}}$ is the large scale channel loss model, and $N$ is the Gaussian noise power.
With $G_{b,k_\mathrm{dB}}$ defined as,
\begin{equation}
    \label{Eqn:LargeScaleChannel}
    \begin{aligned}
    & G_{b,k_\mathrm{dB}}(\rho,\phi,\theta) = \mathrm{PL_{dB}}(\rho) + \mathrm{SF_{dB}}(\rho,\phi,\theta) + A_\mathrm{dB}(\phi,\theta), \\  
    \end{aligned}
\end{equation}
where $\mathrm{PL_{dB}}$ is the path loss, $\mathrm{SF_{dB}}$ is the shadow fading, and $A_\mathrm{dB}$ is the antenna array gain.
% along the horizontal~$\phi$ angle and vertical~$\theta$ angle between the user $k$ and \gls{bs} $b$. 
The path loss and shadow fading models vary depending on the environmental context, ranging from the simplistic \gls{fspl} to the \gls{3gpp} channel models defined in TR 38.901. %~\cite{TR38901}.
These path loss models are generally coupled with shadow fading models that depend on \gls{los} conditions and surrounding terrain information.
The ideal antenna gain patterns can be characterized before deployment. 
However, actual beam patterns may deviate based on the surrounding reflections at the \gls{bs} site. 

Together, these effects are best studied in the spherical coordinate system originating from the \gls{bs}.
This outward radiation phenomenon, characterized by radial path loss attenuation and angular antenna beam pattern correlation, is captured in~(\ref{Eqn:LargeScaleChannel}).
The equation also features a shadow fading component with regional correlation properties that are better captured in the Cartesian coordinate system.
These spatial correlation properties are crucial for interpolation in \gls{rem} generation, and we quantify this understanding using the 3D signal strength dataset presented in~\cite{SungJoonKriging}.

\subsection{Spatial Correlation Properties of 3D Signal Strength Data}

The dataset presented in~\cite{SungJoonKriging} comprises \gls{rsrp} measurements from a \gls{bs} collected at five distinct altitudes, spanning from 30~m to 110~m using an \gls{uav} at the \gls{aerpaw} testbed. 
The authors used this dataset to conduct a correlation analysis of the shadowing component in received signals across space. 
Specifically, Sections VI-A and VI-B of their work detail a joint correlation equation for the shadowing component based on both horizontal and vertical separation distances between pairwise spatial points.
However, rather than separating the contributions of path loss and shadowing, we investigate the correlation of total received signal strength values in 3D space, as follows: 
\begin{enumerate}
    \item We define a spherical grid system with azimuth and elevation angles binned at 0.1~radian increments. Measurement points at varying radii from each 3D angular bin are then grouped. 
    \item From each 3D angular bin group, we compute pairwise \gls{rsrp} correlations and sort the resulting values based on their radial separation into bins of size 5~m. 
    \item The pairwise correlation values within each radial separation bin are then averaged across all 3D angular bins to derive a combined \gls{rsrp} correlation plot as a function of radial separation distance. 
\end{enumerate} 

\begin{figure}
	\includegraphics[width=0.9\linewidth]{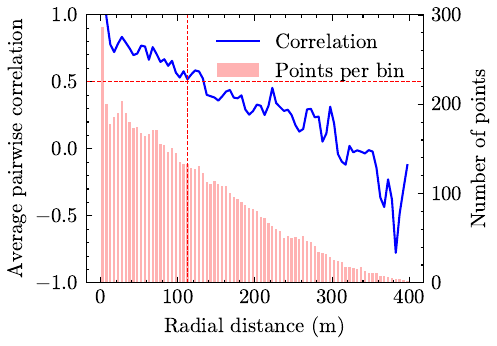}
	\centering
	\caption{Radial distance up to 100~m have correlation values greater than 0.5.}
    \label{fig:SungJoonData}
\end{figure}

We note that the correlation plot in Fig.~\ref{fig:SungJoonData} shows a strong correlation along the radial direction, extending up to a range of 100~m, whereas the shadow fading correlation behavior in~\cite{SungJoonKriging} drops off at less than 50~m.   
This observation aligns with~(\ref{Eqn:LargeScaleChannel}), reinforcing the need to interpolate radio signals in the spherical coordinate system while retaining a Cartesian view for shadow fading interpolation.  
This joint correlation analysis provides a comprehensive understanding of the spatial dependencies of radio signals, thus forming the basis of our proposed \gls{rem} reconstruction technique.

\begin{figure*}
	\includegraphics[width=0.8\linewidth]{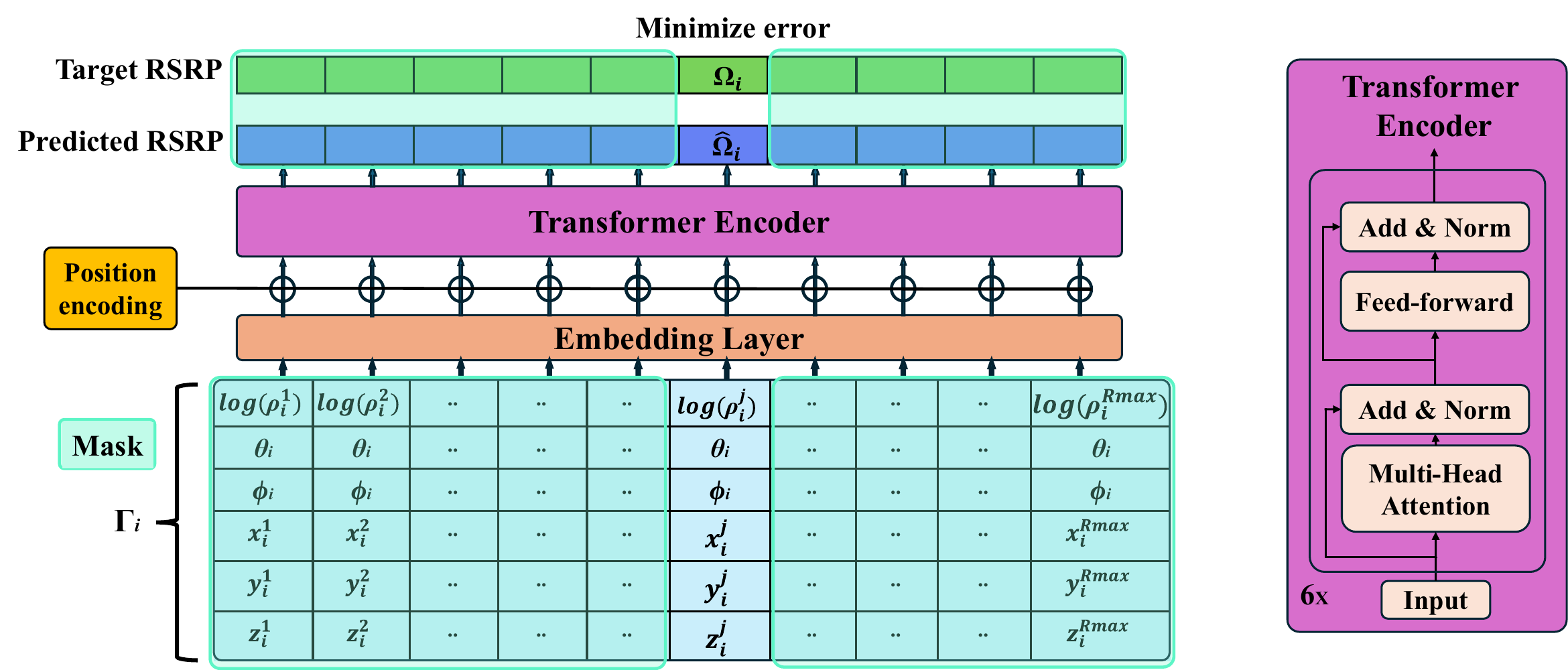}
	\centering
	\caption{The masked input feature $\Gamma_i$ conveys the position information of the $i^\mathrm{th}$ spatial point, and the transformer encoder is trained to predict its \gls{rsrp} $\Omega_i$.}
    \label{fig:TransformerModel}
\end{figure*}

\section{TransfoREM Sytem Design}\label{sec:SystemDesign}

Recognizing the radial nature of radio wave propagation, we formulate the \gls{rem} generation as a sequence estimation task of varying signal strengths originating from the \gls{bs}. 
Fig.~\ref{fig:Spherical_Coord} depicts the radio wave propagation to each point in space as a unique sequence.
Leveraging the strong sequence predictive capabilities of Transformers, we develop the TransfoREM estimator. 
This estimator is initially trained on a deterministic model and then fine-tuned using real-world data to improve prediction accuracy.

\subsection{Radio Environment Viewed in Spherical Coordinate System}
 
Every 3D coordinate [x,y,z] in space is transformed into the spherical coordinate system [$\rho$, $\phi$, $\theta$], as seen in Fig.~\ref{fig:Spherical_Coord}, using the formulae:
\begin{align}
     \mathsf{\rho} &= \sqrt{ x^2 + y^2 + z^2} ~,\\
     \mathsf{\phi} &= \tan^{-1}\frac{y}{x}~,\\
     \mathsf{\theta} &= \cos^{-1}\frac{z}{\rho}.
\end{align}
Here, we set the maximum range of interest to 500~m with a radial step size of 1~m. 
Future adaptations of this method can vary both the maximum range and step size, depending on the context length of the respective applications.
To train a transformer to predict radio propagation behavior, the scene surrounding the \gls{bs} is interpreted as a collection of unique radio propagation sequences.
We then create a range array termed $\delta^{1:R\rm{max}}$, containing ascending steps of radius up to a maximum range $R\rm{max}$. 
Each spatial coordinate is characterized by a distinct $[\phi, \theta]$ 3D angle pair and $\rho$ value, which is binned into a position within $\delta^{1:R\rm{max}}$.
This representation of unique radio propagation directions enables us to construct \gls{rem}s with discrete steps along the radial domain while still retaining continuous angular domain information.  

\subsection{Transformer based \gls{rem} Generation}

Given a dataset $S$ with $N$ 3D spatial points, for each point $i$ $\in$ $[1,...,N]$ with coordinates [$x_i$, $y_i$, $z_i$] and their \gls{rsrp} values $\Omega_i$, we generate a radio propagation feature sequence $\Gamma_i$ that encodes the 3D spatial direction.
$\Gamma_i$ is an array of dimension $6\times R\rm{max}$, including $\big[\log_{10}(\delta^{1:R\rm{max}});\, \theta_i\times I;\, \phi_i\times I;\, x_i^{1:R\rm{max}};\, y_i^{1:R\rm{max}};\, z_i^{1:R\rm{max}}\big]$, where $I$ is an identity vector of dimension $1 \times R\rm{max}$, while $x_i^{1:R\rm{max}}$, $y_i^{1:R\rm{max}}$, and $z_i^{1:R\rm{max}}$ are the cartesian coordinates at every radial bin position along the $i$th point's angular direction. 

The feature vector $\Gamma_i$ is provided as the input feature to an encoder-only transformer model, as depicted in Fig.~\ref{fig:TransformerModel}. 
The encoder model approach is inspired by the foundational wireless transformer model presented in~\cite{largewirelessmodel}.
The input is parsed through an embedding layer, which extracts higher-dimensional features, followed by a position encoding layer, which incorporates their relative position within the sequence. 
Next, the transformer encoder comprises 6 attention layers and 8 attention heads within each multi-head attention block.
Similar to the training of most transformers and large foundational models, we adopt a two-step approach to learn radio wave propagation behavior. 
First, we pre-train the model using synthesized signal strength data, which is generated from the \gls{fspl} model with its characterized antenna gain pattern while excluding any shadowing components. 
Subsequently, we fine-tune the model's predictive performance using real-world data.
This implicitly allows the model to learn the deviated antenna patterns and shadowing components observed in the data, thereby orienting its predictions closer to each BS site's specific radio propagation environment.

\subsubsection{\textbf{Model-based pre-training (Stage-1)}}
\begin{figure}
	\includegraphics[width=\linewidth]{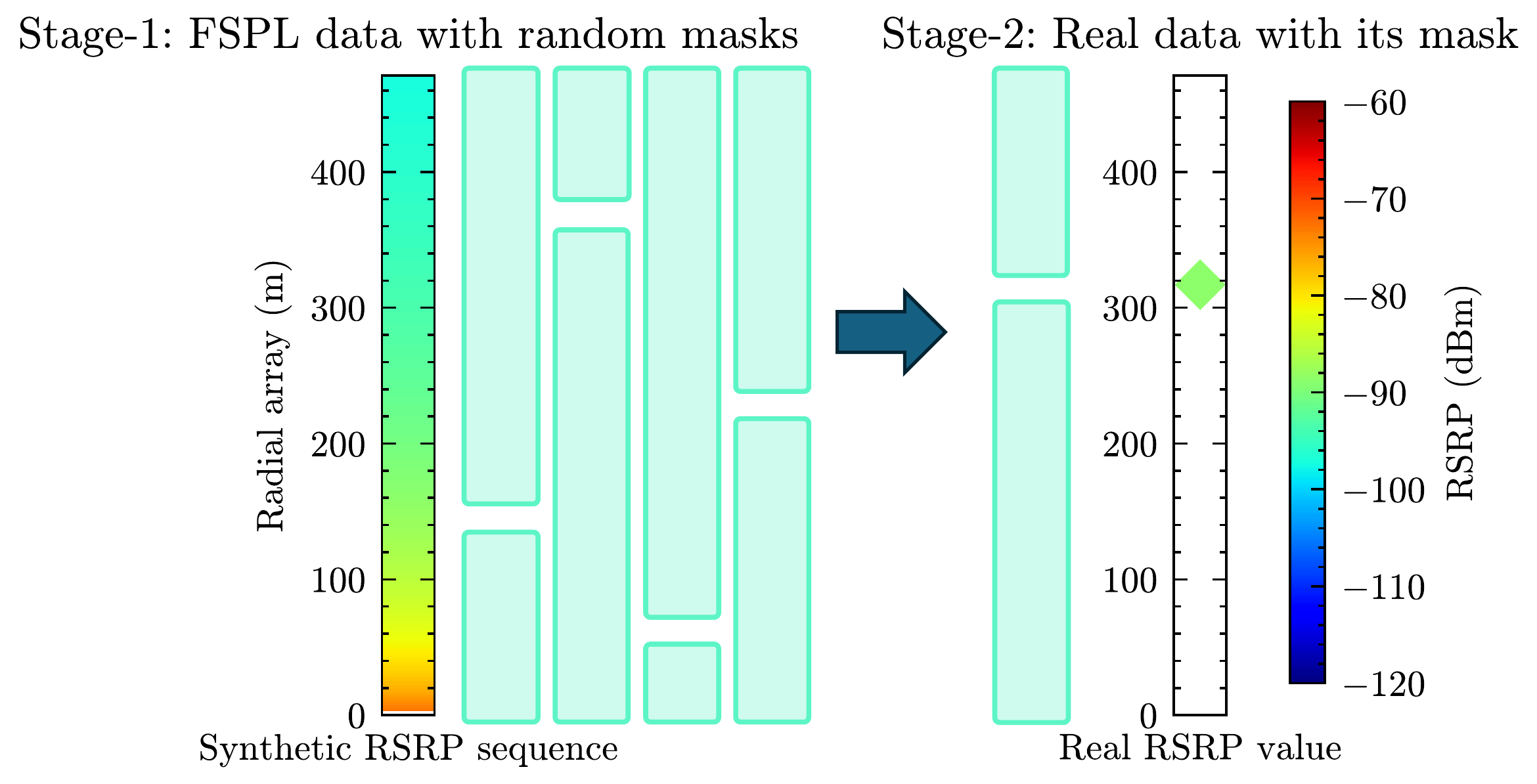}
	\centering
	\caption{The feature mask from Fig.~\ref{fig:TransformerModel} is varied by training stage. The pretraining stage leverages complete sequence information by using random masks at each training iteration, and the fine-tuning stage corrects predictions using masks to align with the limited real-world data.}
    \label{fig:TrainingStageMasking}
\end{figure}

In this step, we generate the \gls{rsrp} sequences for every point $i$ in space, along its angular direction $\phi_i,\theta_i$. The power $P^j_{b,i_\mathrm{dB}}$ at each element $j$ along the $\delta^j_i$ vector is given by,
\begin{equation}
    \begin{aligned}
        P^j_{b,i_\mathrm{dB}}(\rho_i^j, \phi_i, \theta_i) &= P_{b_\mathrm{dB}} + \mathrm{FSPL}_{b,i_\mathrm{dB}}(\rho_i^j) + A_\mathrm{dB}(\phi_i,\theta_i), \\
        \mathrm{FSPL}_{b,i_\mathrm{dB}}(\rho_i^j) &= 147.55 -20\log_{10}(\rho^j_i) -20\log_{10}(f_c),
    \end{aligned}
\end{equation}  
where $P_{b_\mathrm{dB}}$ is the \gls{bs} transmit power, $\mathrm{FSPL}^j_{b,i_\mathrm{dB}}$ is the free space path loss value depending on $\rho^j_i$, the transmit frequency $f_c$, and finally, the antenna directivity gain through $A_\mathrm{dB}$. 
These synthesized \gls{rsrp} vectors are the target vectors used to learn the inverse square law of signal strength attenuation. 
The pre-training stage involves masking the input features and the target vector at random positions along the entire sequence length, inspired by~\cite {largewirelessmodel}.
The availability of a continuous sequence of synthetic data in each propagation direction allows random position masking as seen in Fig.~\ref{fig:TrainingStageMasking}. 
The synthetic dataset from surrounding space is used for batch-training with \gls{mse} loss function, and a \gls{lwsrd} learning rate adaptation algorithm to quickly reach convergence. 
The pretraining stage prepares the model for fine-tuning at each \gls{bs} site based on site-specific real-world data.

\subsubsection{\textbf{Data-driven fine-tuning (Stage-2)}}
In this step, we use real-world data to generate the $\Gamma_i$ feature vectors. 
We generate masks to block the portion of the input features, except for the $k$-th column, where $k$ is the bin corresponding to the position of point $i$ in $\delta^{1:R\rm{max}}$. 
The mask dimensions are $6 \times R\rm{max}-1$, and it serves to convey only relevant features of the $k$-th column to the encoder model. 
The target vector will only contain the real-world \gls{rsrp} value at the $k$-th position and is used to fine-tune predictions accordingly.
The real-world signal strength data points tend to fluctuate drastically due to phase-mismatch and noise issues at the receiver.
To prevent overfitting to such inconsistencies, we adopt the Smooth L1 loss function, which is robust to large deviations during training.
Despite the extensive masking of both input features and outputs necessitated by the limited number of real-world samples, the encoder exhibits robust generalization capabilities.

\section{Evaluation and Results}\label{sec:Results}
In this section, we present the evaluation studies of our two-stage transformer-based \gls{rem} generation technique. 
% We rely on the previously analyzed \gls{aerpaw} dataset and split it into training and testing sets for performance evaluation.
First, we test the accuracy of \gls{rem} reconstruction using data from~\cite{SungJoonKriging} after each training stage. 
With the same dataset, we then compare our results against the Kriging algorithm implementation in~\cite{MushfiqurKriging}, treating it as a baseline method.
Finally, we use a different dataset from~\cite{empirical3dchannelmodeling} and perform further validation of our method. 

\begin{table}
\caption{ Transformer encoder model training parameters.}
\begin{center}
\begin{tabular}{|c|c|c|}
\hline
  Parameter & Stage-1 & Stage-2\\
\hline
  Embedding layer dimension  & \multicolumn{2}{c|}{64}   \\
\hline
  Number of encoder layers   & \multicolumn{2}{c|}{6}   \\
\hline
  Number of attention heads  & \multicolumn{2}{c|}{8}    \\
\hline
  Maximum sequence length    & \multicolumn{2}{c|}{Rmax} \\
\hline
  Loss function              & MSE & Smooth L1 loss \\
\hline
  Learning rate technique    & \gls{lwsrd} & Step decay \\
\hline  
  Learning rate range        & 0.0005-0.0001 & 0.00005-0.00001 \\
% \hline
%   Learning rate warmup steps & 100 & NA \\
\hline
  Batch size                 & 16 & 4\\
\hline
  Epochs                     & 10 & 100\\
\hline
\end{tabular}
\end{center}
\label{Tab:TrainingParams}
\end{table}

\subsection{Two-stage Training Performance}
\begin{figure}
	\includegraphics[width=0.95\linewidth]{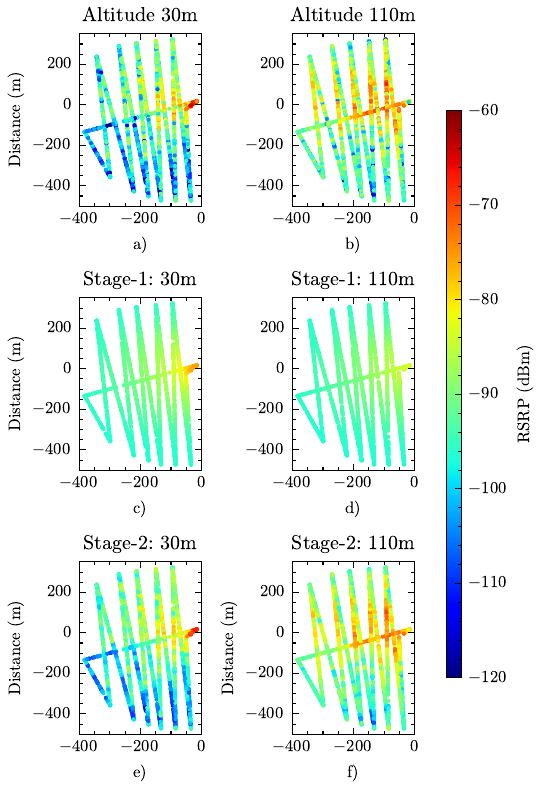}
	\centering
	\caption{a) and b) are the test datasets at the lowest and highest altitude. c) and d) are the stage 1 test set predictions learned from FSPL data. Finally, e) and f) are the best fit test predictions after stage 2 fine-tuning.}
    \label{fig:TwoStageTrainingResults}
\end{figure}

The \gls{aerpaw} dataset~\cite{SungJoonKriging} contains 96,000 points in total.
We perform a filtering step to remove points reporting \gls{rsrp} values below -120 dBm, leaving approximately 17,000 points at each altitude.
We divide the total data into a train:validate:test split with a $0.75:0.05:0.2$  ratio. 
These data points are correspondingly translated into their respective $\Gamma_i$ feature representations. 
Table~\ref{Tab:TrainingParams} contains the training parameters used for the two-stage training used for this evaluation.
% The encoder model is pre-trained using synthetic \gls{fspl} based data and is then fine-tuned in the second stage using real-world data.
Fig.~\ref{fig:TwoStageTrainingResults} depicts the \gls{rem} generation against the test set after each training stage, and the quantitative results are recorded in Table~\ref{Tab:Results}.

This dataset contained a strongly upward-directed antenna beam with increasing energy at higher altitudes of the measurement data.
However, this beam pattern was not evident from the provided antenna pattern. 
As a result, the stage 1 results show weaker signal strength values at higher altitudes, whereas stage 2 fine-tuning refines the \gls{rem} according to the observations from the dataset. 
This behavior is observed in the subplots c) vs e) and d) vs f) in Fig.~\ref{fig:TwoStageTrainingResults}.
We make use of the \gls{rmse} and \gls{mae} metrics for a robust evaluation of regression accuracy.
We also include the coefficient of determination $R^2$, defined as: % $\mathrm{R^2}$,
\begin{equation}
    \begin{aligned}
        R^2 &= 1 - \frac{\sum_{i=1}^{N}( \Omega_i - \hat{\Omega}_i )^2}{\sum_{i=1}^{N}( \Omega_i - \bar{\Omega} )^2}
    \end{aligned}
\end{equation}
where $\Omega_i$ is the real value, $\hat{\Omega}_i$ is the predicted value, and $\bar{\Omega}$ is the average of all real values, to convey the ability to model the variability in the data. 
From Table~\ref{Tab:Results}, we observe that stage 2 fine-tuning achieves improved prediction performance with nearly 3 dB improvement compared to stage 1 in both error metrics and also a desirable increase in $R^2$ value. 

\begin{table}
    \caption{ REM reconstruction results. Metrics for TripleLayerML are extracted from~\cite{empirical3dchannelmodeling} as reported therein.}
    \centering
    \begin{tabular}{ c c c c c }
    \hline
        \textbf{Dataset} & \textbf{REM Method} & \textbf{RMSE} & \textbf{MAE} & $\textbf{R}^2$\\
    \hline
        AERPAW  & TransfoREM Stage 1 & 7.49 dB & 6.20 dB & 0.33 \\
    \cline{2-5}
        dataset~\cite{SungJoonKriging}  & TransfoREM Stage 2 & \textbf{\underline{4.57 dB}} & \textbf{\underline{3.13 dB}} & \textbf{\underline{0.77}}\\
    \hline
        ~  & TransfoREM Stage 1 & 5.47 dB & 3.43 dB & -0.27\\
    \cline{2-5}
        ~ & TransfoREM Stage 2 & \textbf{1.29 dB} & \textbf{0.78 dB} & \textbf{0.93}\\
    \cline{2-5}
         Case Study & TripleLayerML Stage 1& 4.07 dB & 3.04 dB & 0.90\\
    \cline{2-5}
         dataset \cite{empirical3dchannelmodeling} & TripleLayerML Stage 2& 1.27 dB & 0.82 dB & 0.95\\
    \cline{2-5}
         ~ & TripleLayerML Stage 3& \textbf{\underline{1.12 dB}} & \textbf{\underline{0.69 dB}} & \textbf{\underline{0.95}}\\
    \hline
    \end{tabular}
    \label{Tab:Results}
\end{table}

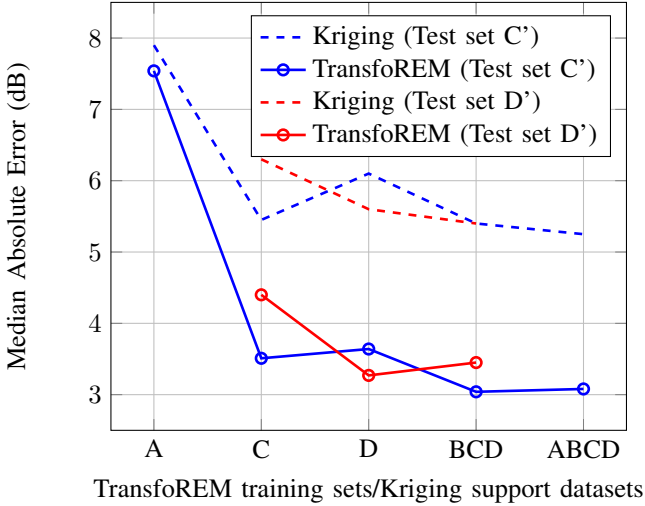
\begin{figure}
\label{fig:AltitudeExtrapolationResults}
\begin{tikzpicture}
    \begin{axis}[
        xlabel={ TransfoREM training sets/Kriging support datasets },
        ylabel={Median Absolute Error (dB)},
        ymin=2.5, ymax=8.5,
        ytick={3, 4, 5, 6, 7, 8},
        grid=major,
        xtick={1,2,3,4,5},
        xticklabels={A, C, D, BCD, ABCD},
        legend pos=north east, % Position the legend in the top-left corner
        legend cell align=left % Align legend text to the left
    ]

    % Main line plot with default markers
    \addplot[
        color=blue,
        dashed, % dashed - Kriging
        line width=1pt
    ] coordinates {
        (1,7.9)     % A
        (2,5.45)    % C
        (3,6.1)     % D
        (4,5.4)     % BCD
        (5,5.25)    % ABCD 
        
    };
    \addlegendentry{Kriging (Test set C')}

    \addplot[
        color=blue,
        mark=o, % circle - TransfoREM
        line width=1pt
    ] coordinates {
        (1,7.54)    % A
        (2,3.51)    % C
        (3,3.64)    % D
        (4,3.04)    % BCD
        (5,3.08)    % ABCD
        
    };
    \addlegendentry{TransfoREM (Test set C')}

    \addplot[
        color=red,
        dashed, % dashed - Kriging
        line width=1pt
    ] coordinates {
                    % A
        (2,6.3)     % C
        (3,5.6)     % D
        (4,5.4)     % BCD
                    % ABCD

    };
    \addlegendentry{Kriging (Test set D')}

    \addplot[
        color=red,
        mark=o, % circle - TransfoREM
        line width=1pt
    ] coordinates {
                    % A   
        (2,4.40)    % C
        (3,3.27)    % D
        (4,3.45)    % BCD
                    % ABCD
        
    };
    \addlegendentry{TransfoREM (Test set D')}

    \end{axis}
\end{tikzpicture}
\caption{TranfoREM with signal propagation sequence learning can extrapolate across altitudes better than shadow correlation-based Kriging interpolation.}
\end{figure}

\subsection{Interpolation and Extrapolation between Various Altitudes}

To build on the \gls{rem} effort in~\cite{MushfiqurKriging}, using their Kriging technique as the baseline, we evaluate \gls{rem} generation using data from different altitudes.
The transformer model in our approach learns radio propagation patterns from the training data through backpropagation and weight updates. 
In contrast, the Kriging method utilizes the semivariogram to select highly correlated data points and performs a covariance-based weighted sum of their signal strengths.
The transformer is a single-shot estimator for each query point, with an inference computational complexity of O($N^2$).
On the other hand, the Kriging technique relies on matrix inversion for weight calculation, which leads to a complexity of O($N^3$).
Overall, the transformer-based technique is well-suited for batch inference and network deployment. 
We borrow the median average error results from~\cite{MushfiqurKriging} and compare the estimation performance between the two techniques.
Following a similar evaluation procedure, we label the data from various altitudes as A:~50~m, B:~70~m, C:~90~m, D:~110~m, and test both 2D and 3D extrapolation properties by learning from various dataset combinations and testing on disjoint subsets C':~90~m and D':~110~m.
Fig. 6 depicts the improved \gls{rem} inference performance of TransfoREM against the kriging technique, with an average improvement of 1.5 dB. 

\subsection{Case Study: Cascaded TripleLayerML architecture} \label{sec:Case Study}
% In this section, we compare the \gls{rem} generation ability of our approach against the work presented in~\cite{empirical3dchannelmodeling}.
The authors of~\cite{empirical3dchannelmodeling} have meticulously collected \gls{lte} \gls{kpis} from a \gls{bs} using a UAV.
% and have published them online for \gls{rem} generation studies.
This published dataset has been filtered to remove outliers and partitioned into 8881 training points and 2144 testing points. 
% This data is very well-conditioned to test the \gls{rem} generation capability.
We utilize this dataset and provide a comparative evaluation against their triple-layer \gls{ml} approach. 
The TripleLayerML approach estimates the \gls{kpis}, and corrects for errors with a new \gls{ml} model at each stage.
The results in Table~\ref{Tab:Results} indicate the stagewise improvement in performance, reaching very low RMSE and MAE values. 
While this approach provides a low REM reconstruction error, training three different \gls{ml} techniques for each \gls{bs} site is more complex, with the third \gls{gpr} layer having O($N^3$) complexity.
The TripleLayerML method also reports a latency-optimized variant that exhibits worse performance, indicating a tradeoff between complexity and accuracy.
From Table~\ref{Tab:Results}, we observe that TransfoREM performs identically to the TripleLayerML after Stage 2. Our approach is also less complex and ready to be deployed at each \gls{bs} for continual online \gls{rem} fine-tuning. 

\section{conclusion}\label{sec:Conclusion}
We frame \gls{rem} generation as a physics-aligned sequence estimation problem and propose a novel two-stage hybrid training approach for a masked Transformer model. 
This method effectively utilizes limited real-world data, demonstrating superior accuracy in REM construction compared to other techniques.
Finally, TransfoREM can be seamlessly integrated into \gls{rem} frameworks in wireless networks to enable the online learning of spatial radio propagation information, which can then be used for informed network management decisions.
 
\bibliographystyle{IEEEtran}
\bibliography{refs}

\end{document}